\documentclass[aps, twocolumn, floatfix, showpacs, nofootinbib,superscriptaddress]{revtex4-1}

\usepackage{graphicx}% Include figure files
\usepackage{dcolumn}% Align table columns on decimal point
\usepackage{bm}% bold math
\usepackage{psfrag}
\usepackage{epsfig}
\usepackage{amsmath}
\usepackage{amssymb}
\usepackage{amsfonts}
\usepackage{color}
\usepackage{mathbbol}
\usepackage{subfigure}
\usepackage{hyperref}

\begin{document}

\title{Bose-glass phases of ultracold atoms due to cavity backaction}
\author{Hessam Habibian}
\affiliation{Theoretische Physik, Universit\"{a}t des Saarlandes, D-66123 Saarbr\"{u}cken, Germany}
\affiliation{Departament de F\'{i}sica, Universitat Aut\`{o}noma de Barcelona, E-08193 Bellaterra, Spain}

\author{Andr\'{e} Winter}
\affiliation{Theoretische Physik, Universit\"{a}t des Saarlandes, D-66123 Saarbr\"{u}cken, Germany}

\author{Simone Paganelli}
\affiliation{Departament de F\'{i}sica, Universitat Aut\`{o}noma de Barcelona, E-08193 Bellaterra, Spain}

\author{Heiko Rieger}
\affiliation{Theoretische Physik, Universit\"{a}t des Saarlandes, D-66123 Saarbr\"{u}cken, Germany}

\author{Giovanna Morigi}
\affiliation{Theoretische Physik, Universit\"{a}t des Saarlandes, D-66123 Saarbr\"{u}cken, Germany}

\date{\today}

\begin{abstract}

We determine the quantum ground-state properties of ultracold bosonic atoms interacting with the mode of a high-finesse resonator. The atoms are confined by an external optical lattice, whose period is incommensurate with the cavity mode wave length, and are driven by a transverse laser, which is resonant with the cavity mode. While for pointlike atoms photon scattering into the cavity is suppressed, for sufficiently strong lasers quantum fluctuations can support the build-up of an intracavity field, which in turn amplifies quantum fluctuations. The dynamics is described by a Bose-Hubbard model where the coefficients due to the cavity field depend on the atomic density at all lattice sites. Quantum Monte Carlo simulations and mean-field calculations show that for large parameter regions cavity backaction forces the atoms into clusters with a checkerboard density distribution. Here, the ground state lacks superfluidity and possesses finite compressibility, typical of a Bose-glass. This system constitutes a novel setting where quantum fluctuations give rise to effects usually associated with disorder. 

\end{abstract}

% 03.75.Hh 	Static properties of condensates; thermodynamical, statistical, and structural properties 
% 05.30.Jp 	Boson systems
% 32.80.Qk 	Phonons or vibrational states in low-dimensional structures and nanoscale materials
% 42.50.Lc 	Quantum fluctuations, quantum noise, and quantum jumps 

\maketitle

Bragg diffraction is a manifestation of the wave-properties of light and a powerful probe of the microscopic structure of a medium: Bragg peaks are intrinsically related to the existence of spatial order of the scatterers composing a medium and provide a criterion for the existence of long-range order \cite{Crystal}. Bragg diffraction of light by atoms in optical lattices has been measured for various geometries and settings, from gratings of laser-cooled atoms \cite{Phillips,Hemmerich,Courteille,Courteille-disorder} to ultracold bosons in the Mott-Insulator (MI) phase \cite{Bloch}. In most of these setups the backaction of light on the atomic medium, due to the mechanical effects of atom-photon interactions, is usually negligible, while photon recoil can give rise to visible effects in the spectrum of the diffracted light \cite{Rist}. 

Recent work proposed to use high-finesse optical resonators to enhance light scattering into one spatial direction, increasing the collection efficiency and thereby suppressing diffusion related to photon scattering \cite{MekhovPRL}. For appropriate geometries, properties of the medium's quantum state can be revealed by measuring the light at the cavity output \cite{Rist,MekhovPRL}. These proposals assume that backaction of the cavity field on the atoms can be discarded. Such an assumption is, however, not valid in the regime considered in Refs. \cite{Domokos2002,Black2003,Nagy2008,Keeling2010,Baumann2010,Baumann2011}: Here, the strong coupling between cavity and atoms can induce the formation of stable Bragg gratings in cold \cite{Domokos2002,Black2003} and ultracold atomic gases \cite{Nagy2008,Keeling2010,Baumann2010,Baumann2011} that coherently scatter light from a transverse laser into the cavity mode. This phenomenon occurs when the intensity of the pump exceeds a certain threshold \cite{Black2003,Asboth2005,Nagy2008,Baumann2010}. At ultralow temperatures the self-organized medium is a supersolid \cite{Baumann2010}, while for larger pump intensities incompressible phases are expected \cite{Fernandez2010}. 

\begin{figure}
\centering
\includegraphics[width=5cm]{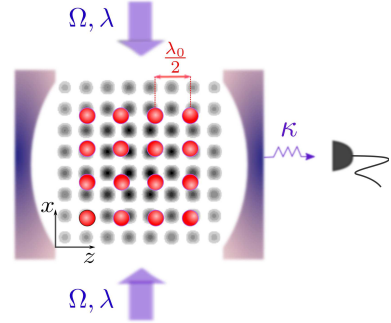}
\caption[]{\label{fig:1}(color online) Ultracold atoms are confined by an optical lattice of periodicity $\lambda_0/2$. They are transversally driven by a laser at Rabi frequency $\Omega$ and strongly coupled to the standing-wave mode of a high-finesse cavity with wavelength $\lambda$. The wave lengths $\lambda$ and $\lambda_0$ are incommensurate: The light scattered into the cavity mode gives rise to an incommensurate potential, which mediates an effective long-range interaction between the atoms and can substantially modify the quantum ground state. }
\end{figure}

Let us now assume that the atoms are inside a high-finesse standing-wave cavity and form a periodic structure, like the one sketched in Fig. \ref{fig:1}, whose period is incommensurate with the cavity mode wavelength. If the atoms are pointlike scatterers, or deep in a MI phase of the external potential, there is no coherent scattering from a transverse laser into the cavity mode \cite{Habibian2011,MekhovPRL}. Quantum fluctuations, however, will induce scattering into the cavity mode, thus the creation of a weak periodic potential which is incommensurate with the periodicity of the optical lattice confining the atoms. In this Letter we derive a Bose-Hubbard model for the system in Fig. \ref{fig:1} and show that cavity backaction gives rise to an additional term in the atomic energy which depends on the density at all lattice sites. Even for weak probe fields this results in the formation of patterns which maximize scattering into the cavity mode and can exhibit finite compressibility with no long-range coherence. This feature, typical of disordered systems, corresponds to a Bose-glass (BG) phase for sufficiently deep potentials \cite{Fisher,BoseGlass,Damski,Deissler2010} and here emerges due to the nonlocal quantum potential of the cavity field. These predictions should be observable in existing experimental setups \cite{Baumann2010,Barrett2012}. These findings extend studies where glassiness was predicted in multimode cavities \cite{Goldbart2010,Goldbart2011}. 

To analyze the quantum dynamics of the atom-cavity system we assume that the atomic motion is confined to the $x-z$ plane where ${\bf r}=(x,z)$ is the atomic position. It is furthermore assumed that the scattering processes are coherent, namely, the modulus of the detuning $\Delta_a=\omega_L-\omega_0$ between the frequencies of pump, $\omega_L$, and atomic transition, $\omega_0$, is much larger than (i) the transition linewidth, (ii) the strength of the atom-photon coupling, and (iii) the modulus of the detuning $\delta_c=\omega_L-\omega_c$ between pump and cavity mode \cite{Baumann2010,Fernandez2010}. The laser and cavity mode have wave vector ${\bf k}_L=k\hat{x}$ and ${\bf k}_{\rm cav}=k\hat{z}$, respectively: When the polarizations are suitably chosen, only one stable electronic state is occupied at all times. We denote by $\hat a$ and $\hat a^{\dagger}$ the operators annihilating and creating a cavity photon and by $\hat\psi({\bf r})$ the bosonic field operator for an atom at ${\bf r}$, with $[\hat\psi({\bf r}),\hat\psi^\dag({\bf r'})]=\delta^{(2)}({\bf r}-{\bf r'})$. In the reference frame rotating at frequency $\omega_L$ the Hamiltonian governing the coherent dynamics reads $\hat H=-\hbar\delta_c \hat a^\dag \hat a+\hat H_a+\hat H_i$, with $\hat H_{\ell=a,i}=\int{\rm d}{\bf r}\hat\psi^{\dagger}({\bf r})\hat{\cal H}_\ell({\bf r})\hat\psi({\bf r})$. Here, $\hat{\cal H}_a({\bf r})$ describes the atomic dynamics in absence of the resonator:
\begin{eqnarray*}
\hat{\cal H}_a({\bf r})=-\frac{\hbar^2}{2m}\nabla^2+V_{0}({\bf r})+{\mathcal G}_{s}\,\hat n({\bf r})+V_1\cos^2(kx)\,,
\end{eqnarray*}
where $V_{0}({\bf r})=V_t\{\cos^2(k_0 z)+\beta\cos^2(k_0 x)\}$ is an external periodic potential of wave number $k_0$, depth $V_t$ along $z$ and aspect ratio $\beta$, $\hat n({\bf r})=\hat\psi^{\dagger}({\bf r})\hat\psi({\bf r})$ is the atomic density operator, ${\mathcal G}_{s}$ is the strength of $s$-wave collisions, and $V_1=\hbar\Omega^2/\Delta_a$ such that $|V_1|$ is the depth of the dynamical Stark shift induced by the transverse standing-wave laser at Rabi frequency $\Omega$. The coupling with the cavity gives $\hat{\cal H}_i({\bf r})$ with \cite{Fernandez2010,Maschler} 
\begin{align}
\label{H:i}
\hat{\cal H}_i({\bf r})=\hbar U_0\cos^2(kz)\hat a^\dag \hat a+\hbar S_0\cos(kx)\cos(kz)\left(\hat a^\dag+\hat a\right)\,,
\end{align}
where $U_0=g_0^2/\Delta_a$ is the dynamical Stark shift per cavity photon and $g_0$ is the vacuum Rabi frequency at a cavity-field maximum. The frequency $S_0=g_0\Omega/\Delta_a$ is the amplitude of scattering a laser photon into the cavity mode by one atom. The corresponding term describes the coherent pump on the cavity field via scattering by the atoms and depends on the atomic positions within the field. It gives rise to a large intracavity-photon number $n_{\rm cav}$ when the atoms form a Bragg grating with periodicity $2\pi/k$. This would correspond to choosing $k=k_0$ \cite{Footnote:1}: $n_{\rm cav}$ would then depend on the balance between the superradiant scattering strength, $S_0N$, and the rate of cavity loss $\kappa$. On the contrary, in this paper we take $k$ and $k_0$ to be incommensurate and analyse the effect of the strong coupling with the cavity field when the atoms are tightly confined by the potential $V_0({\bf r})$ and weakly pumped by the transverse laser. 

\begin{figure*}
\centering
\includegraphics[width=5.9cm]{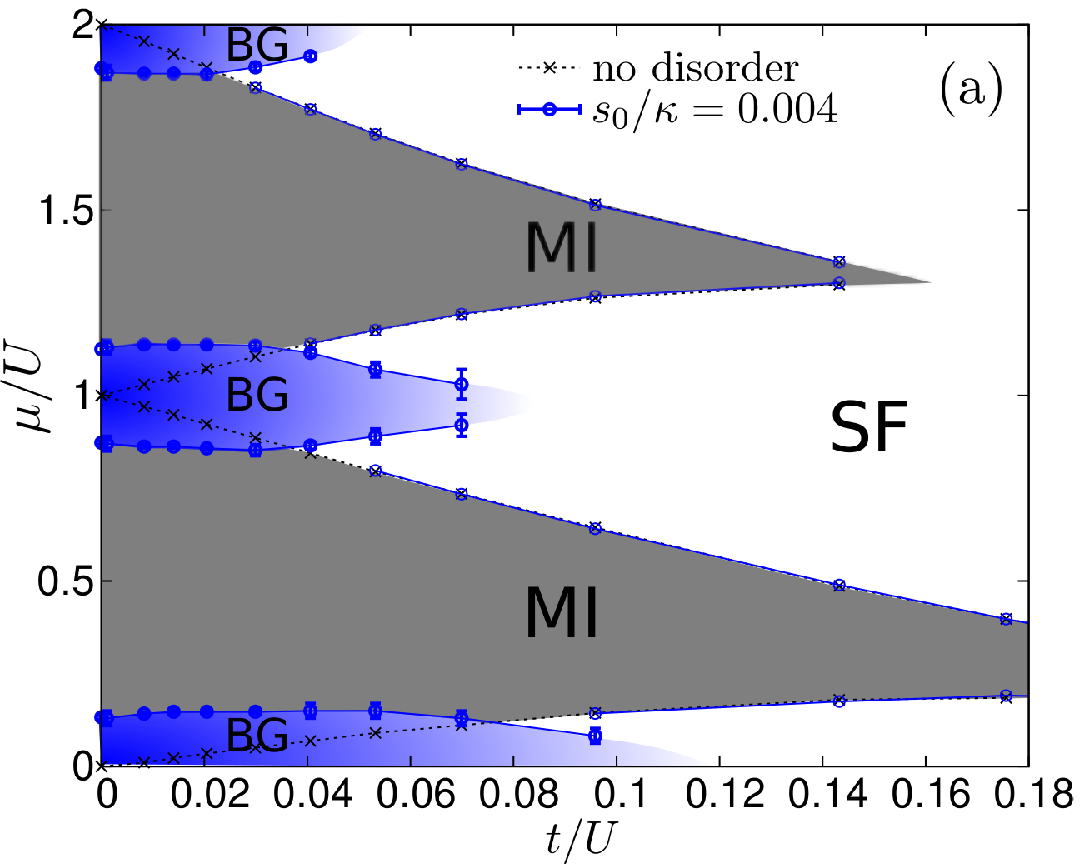}
\includegraphics[width=5.9cm]{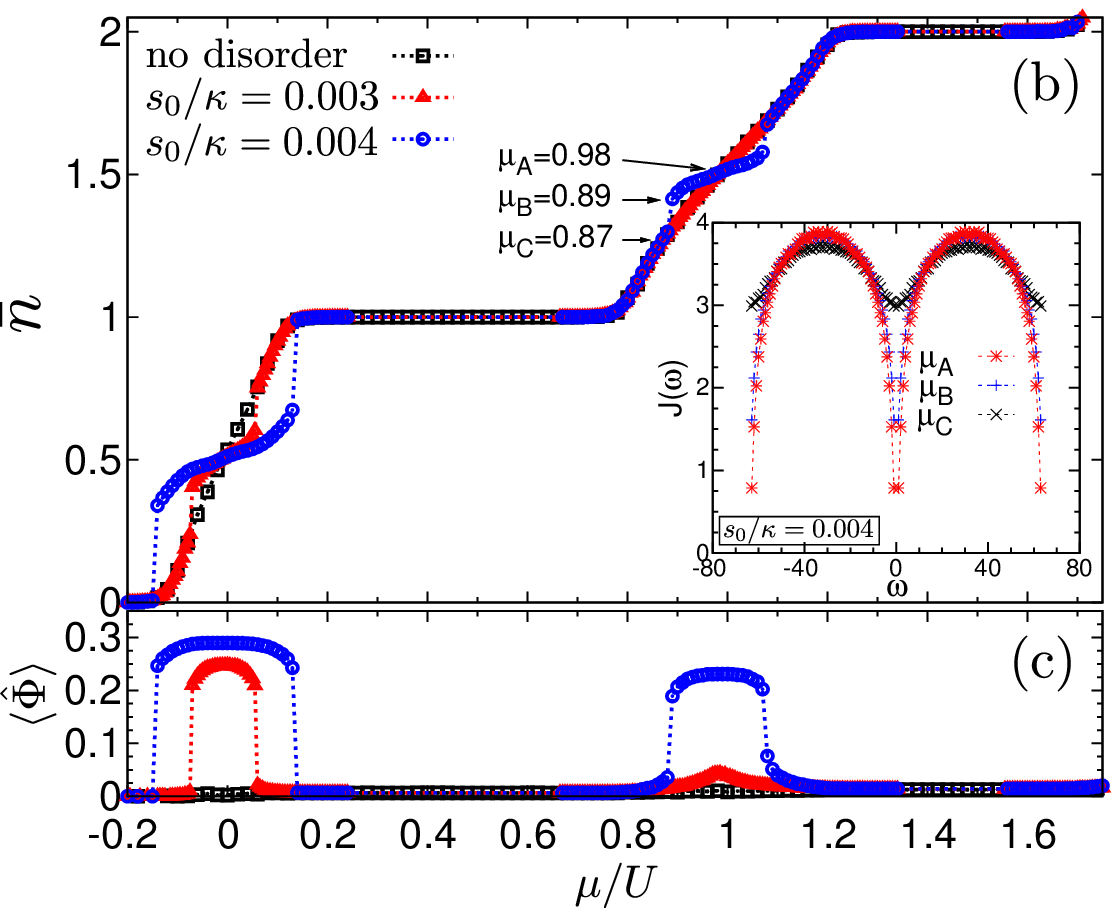}
\includegraphics[width=5.9cm]{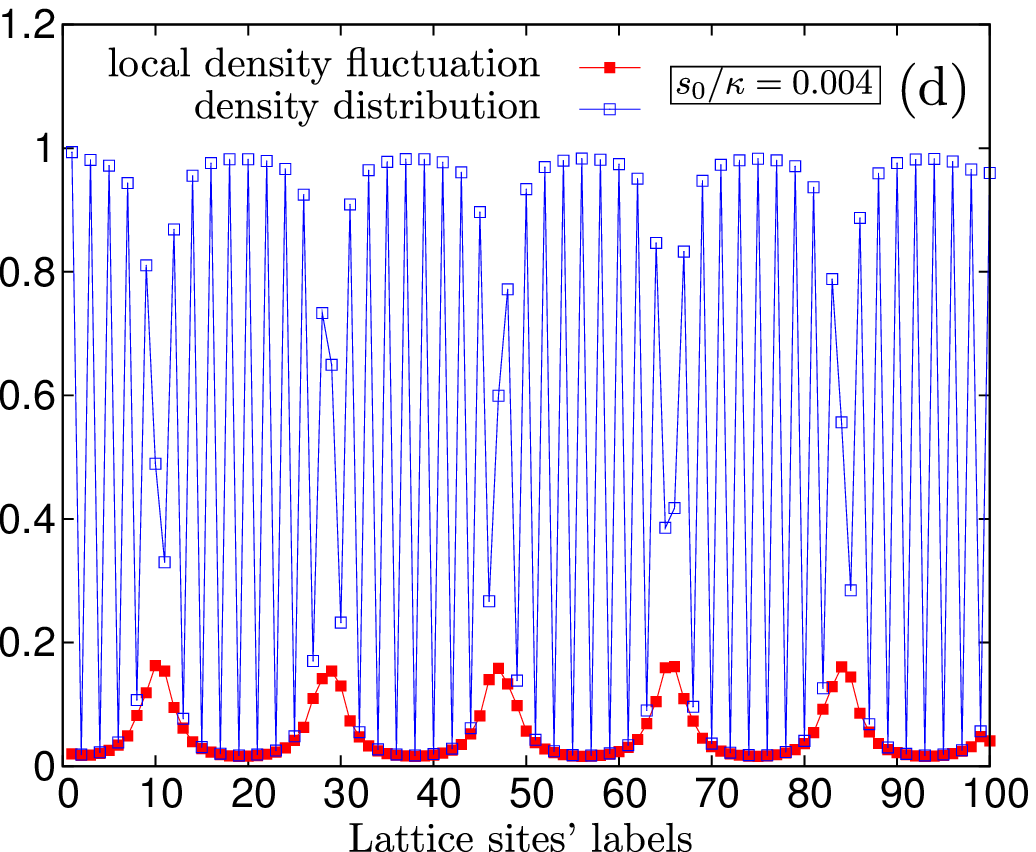}
\caption{ \label{Fig:MC}(color online) Results of QMC simulations for a 1D lattice ($\beta\gg1$) with 100 sites and periodic boundary conditions. (a) Phase diagram in the $\mu-t$ plane for $s_0=0.004\kappa$. The gray regions indicate incompressible phases at density $\bar{n}=1,2$, the blue region the gapless phases with vanishing SF density. The dotted lines indicate the shape of the incompressible phases in absence of the pump laser ($s_0=0$) \cite{Footnote:Batrouni}. (b) Linear density $\bar{n}$ and (c) $\langle\hat\Phi\rangle$ as a function of $\mu$ for $t=0.053 U$ and $s_0/\kappa=0.003$ (triangles), $s_0/\kappa=0.004$ (circles), and $s_0=0$ (squares): The number of photons is different from zero for the parameters corresponding to the blue regions in (a). The inset of (b) displays the Fourier transform of the pseudo current-current correlation function $J(\omega)$ \cite{Footnote:Batrouni} for the parameters indicated by the arrows in the curve of (b). (d) Local density distribution $\langle \hat{n}_i\rangle$ (empty points joined by the blue curve) and local density fluctuations $\langle\hat{n}_i^2\rangle-\langle\hat{n}_i\rangle^2$ (filled points joined by the red curve) as a function of the site for $\mu=0$ and $s_0=0.004\kappa$. The values of $s_0$ are found from $s_0=\sqrt{K}\Omega g_0/\Delta_a$ and are consistent with the parameters of Ref. \cite{Baumann2010} ($g_0/2\pi=14.1$ MHz, $\Delta_a/2\pi=58$ GHz, $\kappa/2\pi=1.3$ MHz). The other parameters are $\delta_c=-5\kappa$, $k/k_0=785/830$ as in \cite{Baumann2010}, while ${\mathcal G}_s$ has been taken from Ref.~\cite{Volz_03}.}
\end{figure*}

We now sketch the derivation of an effective Bose-Hubbard Hamiltonian describing the dynamics of the atoms in the cavity potential. We first assume that in the time scale $\Delta t$ the atomic motion does not significantly evolve while the cavity field has reached a local steady state, $\int_t^{t+\Delta t}\hat a(\tau)d\tau/\Delta t\approx \hat a_{\rm st}$. Here,  $\int_t^{t+\Delta t}\dot{\hat{a}}_{\rm st}(\tau)d\tau=0$ where the time evolution is governed by the Heisenberg-Langevin equation of motion $\dot{\hat{a}}= [\hat a,\hat H]/{\rm i}\hbar-\kappa \hat a+\sqrt{2\kappa}\hat a_{\rm in}(t)$, with  $\hat a_{\rm in}(t)$ the input noise such that $[\hat a_{\rm in}(t),\hat a^\dag_{\rm in}(t')]=\delta(t-t')$. The conditions for a time scale separation are set by the inequality $|\delta_c+{\rm i}\kappa|\Delta t\gg 1$ and by the assumption that coupling strength between atoms and fields are much smaller than $1/\Delta t$ \cite{Footnote:Scales}. The stationary field reads
\begin{equation}\label{a_mf}
\hat a_{\rm st}=\frac{S_0\hat{\mathcal Z}}{(\delta_c-U_0\hat{\mathcal Y})+{\rm i}\kappa}+\frac{{\rm i}\sqrt{2\kappa}\hat{\bar{a}}_{\rm in}}{(\delta_c-U_0\hat{\mathcal Y})+{\rm i}\kappa}\,,
%\left[\int_{t_0}^{t_0+\Delta t} a_{in}(\tau){\rm d}\tau\right]/\Delta t\nonumber
\end{equation}
with $\hat{\bar{a}}_{\rm in}$ the input noise averaged over $\Delta t$. Here, $\hat{\mathcal Y}=\int {\rm d}^2{\bf r}\cos^2(kz)\hat n({\bf r})$ and  $\hat{\mathcal Z}=\int {\rm d}^2{\bf r} \cos(kz)\cos(kx)\hat n({\bf r})$  describe the shift of the cavity resonance and coherent scattering amplitude, respectively, due to the atomic density distributions. The quantum noise term can be neglected when the mean intracavity-photon number is larger than its fluctuations, that corresponds to taking $|S_0\langle\hat{\mathcal{Z}}\rangle|\gg\kappa$. In this limit, the field at the cavity output, $\hat a_{\rm out}=\sqrt{2\kappa}\hat a_{\rm st}-\hat a_{\rm in}$, allows one to non-destructively monitoring the state of the atoms \cite{Habibian2011}.
Using Eq. \eqref{a_mf} in place of the field $\hat a$ in Eq. \eqref{H:i} leads to the effective atomic Hamiltonian, from which we derive a Bose-Hubbard model assuming that the atoms are tightly trapped in the lowest band of $V_0({\bf r})$. We use the Wannier decomposition $\hat \psi({\bf r})=\sum_i w_i({\bf r})\,\hat b_i$, with  $w_i(\bf r)$ the lowest-band Wannier function centered at the minimum $\bf r_i$ of the classical potential $V_0({\bf r})$ and $\hat b_i$ the bosonic operator annihilating a particle at  $\bf r_i$, such that  $\hat{n}_i=\hat b_i^{\dagger}\hat b_i$ is the onsite atomic density. Using the thermodynamic limit where the cavity parameters scale with the number of lattice sites $K$ according to $S_0=s_0/\sqrt{K}$ and $U_0=u_0/K$ \cite{Fernandez2010}, the resulting Bose-Hubbard Hamiltonian reads
\begin{equation}\label{H_BH}
\hat{\mathcal H}_{{\rm BH}}=\sum_{i=1}^K\left( -\sum_{\langle j,i\rangle} \hat t_i(\hat b_i^\dag \hat b_{j}+{\rm H.c.})+\frac{U}{2} \hat{n}_i(\hat{n}_i-1)-\hat \mu_i\hat{n}_{i}\right),
\end{equation}
with $U={\mathcal G}_{s}\int d^2{\bf r}\, [w_i({\bf r})]^4$ the onsite interaction strength, $\hat\mu_i$ the site-dependent chemical potential, $\hat t_i$ the tunneling coefficient, and $\langle j,i\rangle$ the sum over nearest neighbors. Here, $\hat \mu_i=\mu^{(0)}+\delta\hat\mu_{i}$, $\hat t_i=t^{(0)}+\delta \hat t_i$, where $\mu^{(0)}=-E_0-V_{t}X_0$ and $t^{(0)}=-E_1-V_{t}X_1$ are constant over the lattice, with $E_{s=0,1}=-\hbar^2/(2m)\int {\rm d}^2{\bf r} \,w_i({\bf r}) \nabla^2w_{i+s}({\bf r})$ and $X_{s=0,1}=\int {\rm d}^2{\bf r} \,\{\cos^2(k_0z)+\beta \cos^2(k_0x)\}w_i({\bf r})w_{i+s}({\bf r})$. The terms $\delta\hat\mu_{i}$ and $\delta \hat t_i$ are due to the pump and cavity incommensurate potentials and vanish when the pump laser is off, $\Omega=0$.  In this limit the model reduces to the regular Bose-Hubbard model exhibiting a MI-superfluid (SF) transition as a function of the parameters $\mu$ and $t$ \cite{Fisher}. When $\Omega\neq 0$, instead, one has
\begin{eqnarray}
\label{delta:mu}
\delta\hat\mu_{i}=-V_1 J_{0}^{(i)}-\hbar\frac{s_0^2}{\hat{\delta}_{\rm eff}^2+\kappa^2}\hat{\Phi}\left( 2\hat{\delta}_{\rm eff}Z_{0}^{(i)}+u_0\hat{\Phi} Y_0^{(i)}\right),\quad\label{mu_i}
\end{eqnarray}
while $\delta\hat t_i$ is negligible, hence $t_i=t^{(0)}$ \cite{Footnote:2}. In Eq. \eqref{delta:mu} we introduced $J_0^{(i)}=\int d^2{\bf r}\cos^2(kx)w_i({\bf r})^2$, which scales the strength of the classical transverse potential due to the laser (it is constant for the sites with same $x$ value). The coefficients for the cavity field read $Z_0^{(i)}=\int d^2{\bf r}\cos(kx)\cos(kz) w_i({\bf r})^2$, $Y_0^{(i)}=\int d^2{\bf r}\cos^2(kz) w_i({\bf r})^2$, while operator $\hat{\delta}_{\rm eff}=\delta_c-u_0\sum_i Y_0^{(i)} \hat{n}_i /K$ accounts for the cavity-frequency shift due to the atoms \cite{Stamper-Kurn,Maschler}. Remarkably, the cavity effects are scaled by the operator
\begin{equation}
\hat{\Phi}=\sum_{i=1}^K Z_0^{(i)}\hat{n}_i/K\,,
\end{equation}
which originates from the long-range interaction mediated by the cavity field, and is related to the mean intracavity photon number since $n_{\rm cav}\propto\langle\hat{\Phi}^2\rangle$. Its mean value vanishes when the atomic gas forms a MI state: $\langle \hat\Phi\rangle_{\rm MI} \propto \sum_iZ_0^ {(i)}=0$, since there is no coherent scattering into the cavity mode. Also deep in the SF phase $\langle \hat\Phi\rangle_{\rm SF} \to 0$. Close to the MI-SF phase transition, however, fluctuations in the atomic density lead to finite values of $\langle \hat\Phi\rangle$, hence to a finite intracavity photon number. The dependence of the chemical potential on the operator $\hat\Phi$ is  a peculiar property of our model, that distinguishes it from the case of a bichromatic optical lattice \cite{Damski,Deissler2010}, where the strength of the incommensurate potential is independent of the phase of the ultracold atomic gas.

\begin{figure*}
\centering
\includegraphics[width=6.3cm]{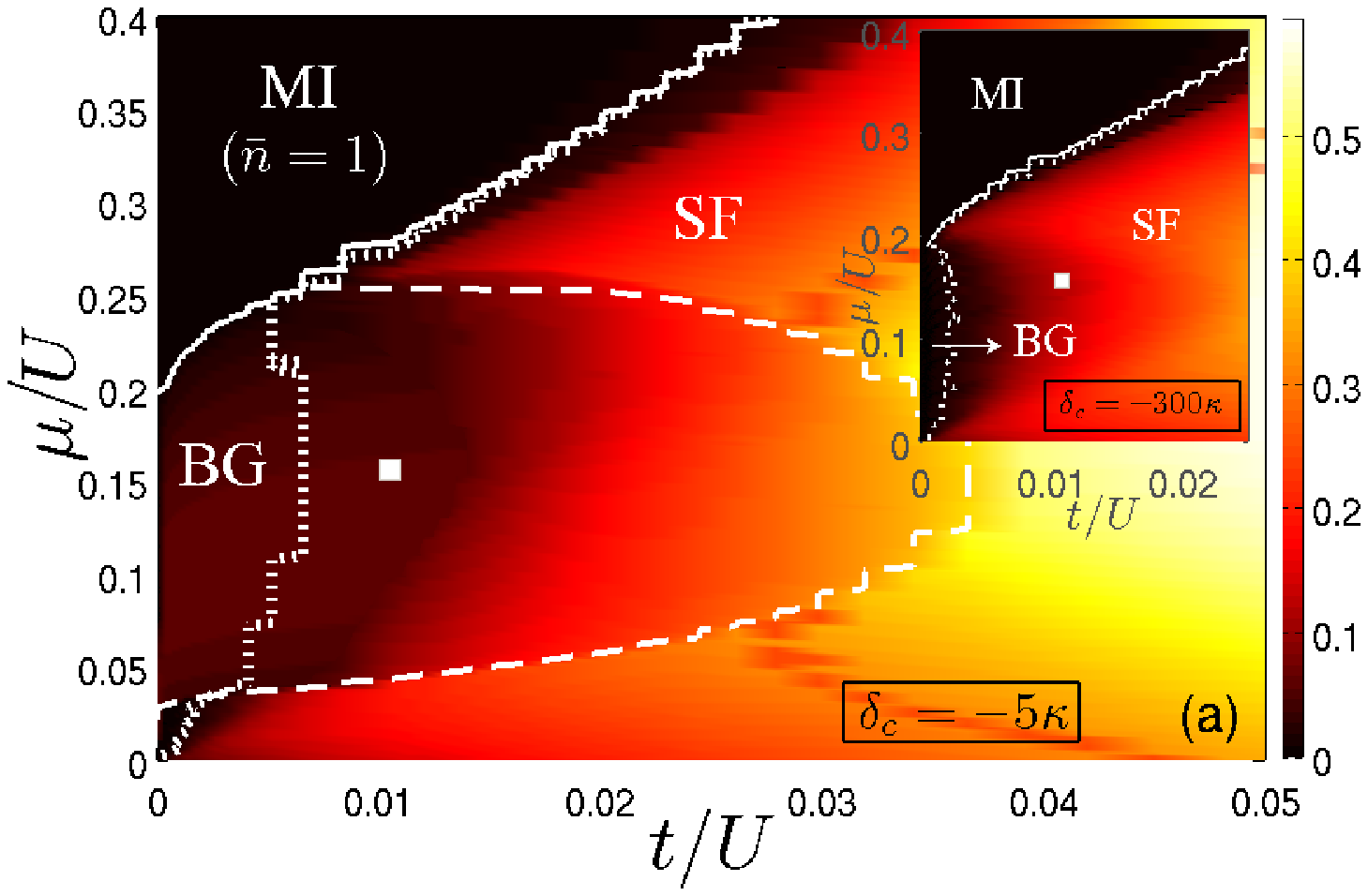}
\includegraphics[width=5.1cm]{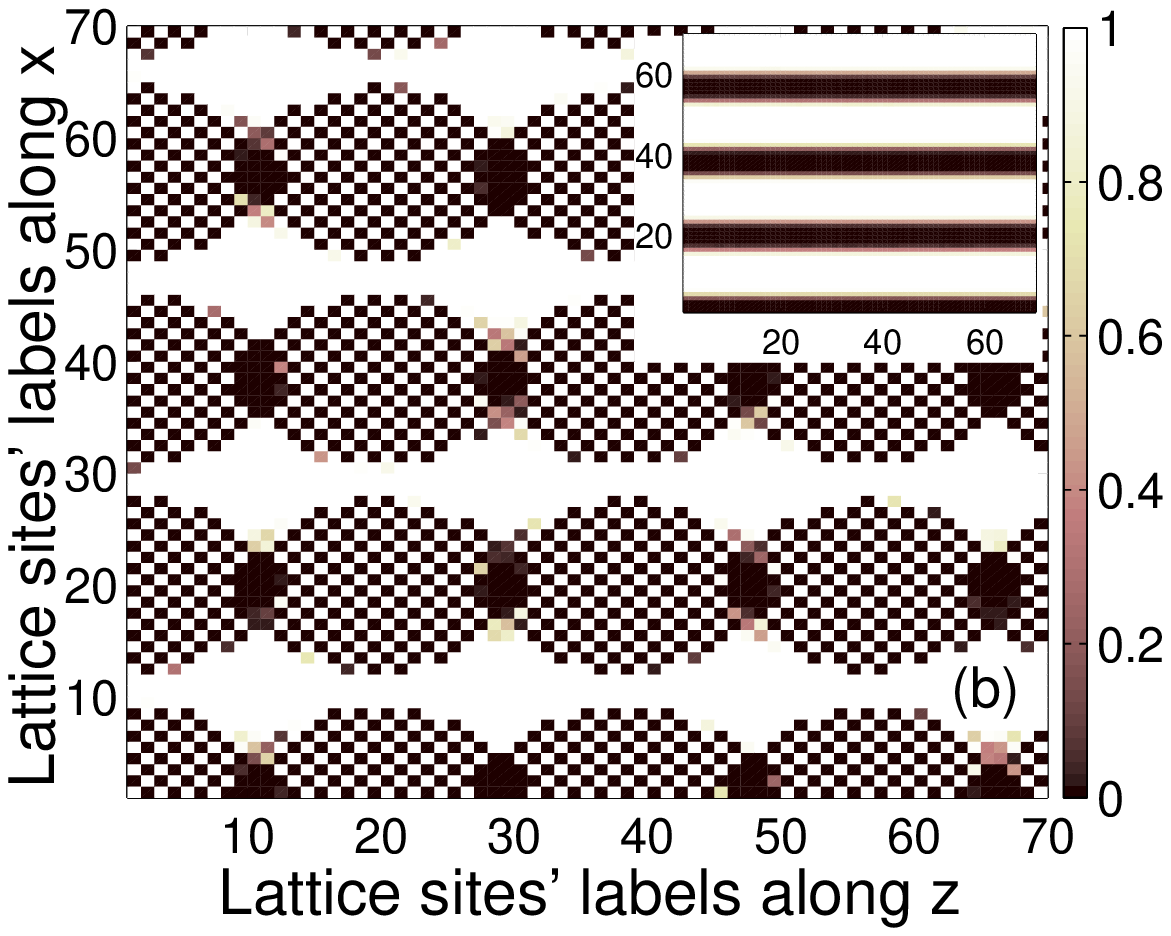}
\includegraphics[width=6.3cm]{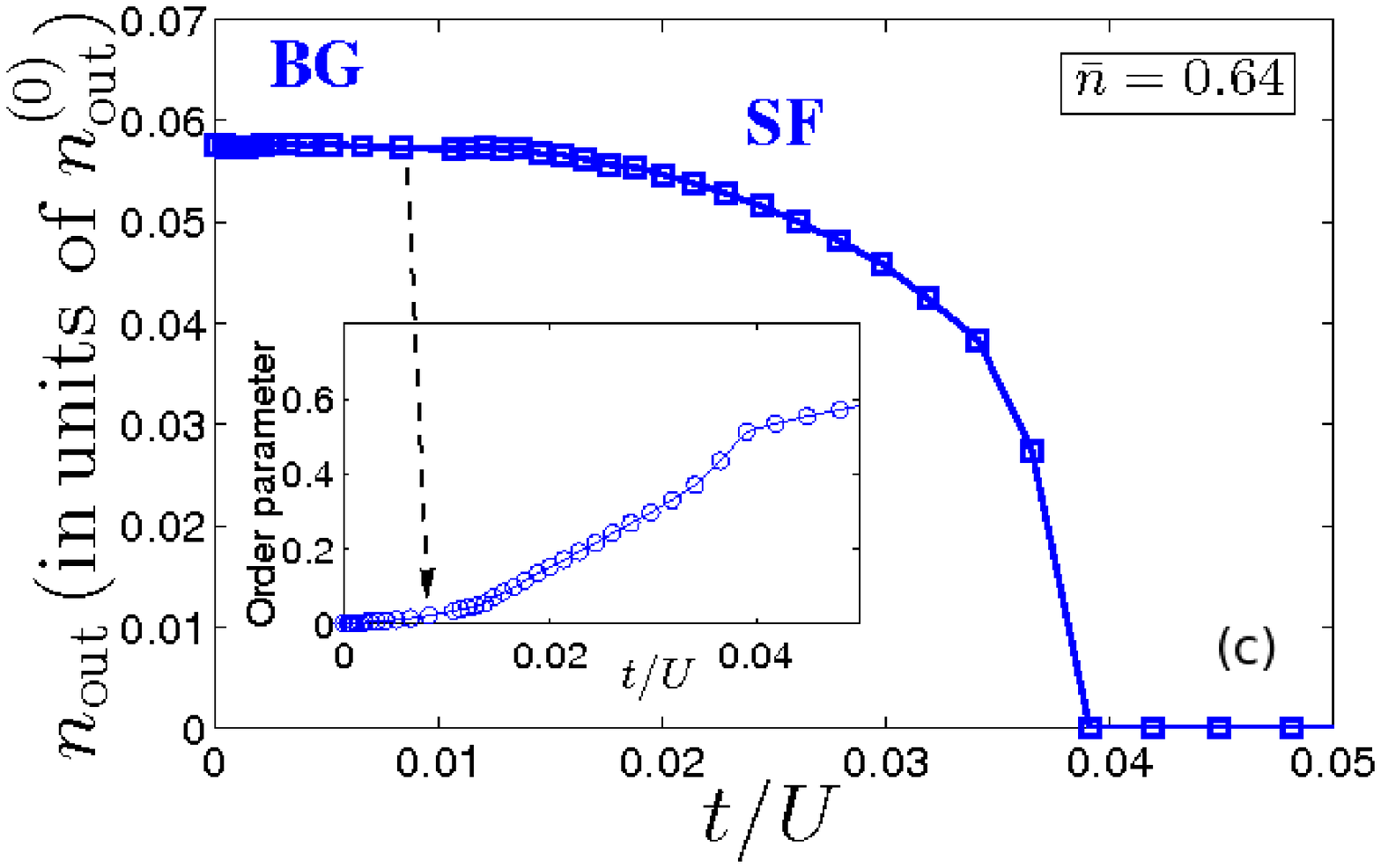}
\caption[]{(color online) Results of a mean-field calculation for a 2D lattice
 with $70\times 70$ sites and periodic boundary conditions. (a) Order parameter in the $\mu-t$ plane when $s_0=0.15\kappa$ ($\delta_c=-5\kappa$). Inset: phase diagram when cavity backaction is negligible ($\delta_c=-300\kappa$) \cite{Footnote:Defect}. The dotted (solid) curve delimitates the regions with a small, arbitrarily chosen threshold for the order parameter (density fluctuations) \cite{Footnote:MFthreshold}. The dashed curve shows the region where the number of photons is 2 orders of magnitude larger than outside. (b) Local density distribution $\langle\hat{n}_i\rangle$ at $\mu=0.156 U$ and $t=0.01 U$ (point in (a) at $\bar{n}=0.64$). Inset: corresponding density distribution in the $\mu-t$ plane when cavity backaction is negligible (here $\bar{n}=0.625$). (c) Photon emission rate at the cavity output, $n_{\rm out}=\langle a_{\rm out}^\dag a_{\rm out}\rangle$, as a function of $t$ for $\bar{n}=0.64$. The rate is in units of $n_{\rm out}^{(0)}=\kappa n_{\rm cav}^ {\rm (max)}$, where $n_{\rm cav}^ {\rm (max)}=s_0^2K/(\hat\delta_{\rm eff}^2+\kappa^2)$ is the maximum number of intracavity photons when all atoms scatter in phase into the cavity mode. Inset: corresponding order parameter. Note that in absence of cavity backaction ($\delta_c=-300\kappa$) the rate is less than six orders of magnitude smaller. The value $s_0=0.15\kappa$ is consistent with the parameters of Ref. \cite{Baumann2010} when taking $300\times 300$ lattice sites. The onsite repulsion ${\mathcal G}_s$ is taken from Ref.~\cite{Kruger_07}.
} \label{Fig:PhaseDiag}
\end{figure*}
We first analyse the ground state of the Bose-Hubbard Hamiltonian, Eq. \eqref{H_BH}, when the system can be reduced to a one-dimensional (1D) lattice along the cavity axis ($\beta\gg1$). In 1D the effect of the transverse potential (first term in Eq. \eqref{delta:mu}) is a constant shift which can be reabsorbed in the chemical potential. Figure~\ref{Fig:MC}(a) displays the phase diagram in the $\mu-t$ plane evaluated by means of a quantum Monte Carlo (QMC) calculation  \cite{Footnote:Batrouni}. The compressibility is determined using $\chi=\frac{\partial{\bar n}}{\partial \mu}$, with $\bar n=\sum_i \langle\hat{n}_i\rangle/K$ the mean density, while the SF density is obtained by extrapolating the Fourier transform of the pseudo current-current correlation function $J(\omega)$ \cite{Footnote:Batrouni} to zero frequency (see inset of panel (b)). The grey regions indicate the MI states at densities $\bar n=1,2$, the blue regions a compressible phase where the SF density vanishes, while outside the phase is SF. The effect of cavity backaction is evident at low tunneling, where $\langle \Phi\rangle >0$: Here the size of the MI regions is reduced. At larger tunneling a direct MI-SF transition occurs and the MI-SF phase boundary merges with the one found for $s_0=0$: In fact, for larger quantum fluctuations $\langle\hat\Phi\rangle\to 0$. This feature is strikingly different from the situation in which the incommensurate potential is classical \cite{Roux08,Deng08}: There, the MI lobes shrink at all values of $t$ with respect to the pure case. The atomic density and corresponding value of  $\langle\hat\Phi\rangle$ are displayed in Fig.~\ref{Fig:MC}(b) and (c) as a function of $\mu$ for different values of the transverse laser intensity (thus $s_0$): the incommensurate potential builds in the BG phase of the diagram, where the number of intracavity photons does not vanish. Panel (d) displays the local density distribution $\langle\hat{n}_i\rangle$ and the local density fluctuations in the BG phase: the density oscillates in a quasi-periodic way over clusters in which the atoms scatter in phase into the cavity mode. The fluctuations are larger at the points where $Z_0^{(i)}$, which oscillates at the cavity mode wave length, becomes out of phase with the trapping potential. In this way the density distribution maximizes scattering into the cavity mode. 

The two-dimensional (2D) case is studied by means of a mean-field calculation \cite{Fisher,Sheshadri} for a pumping strength $s_0$ such that $\max\{|\langle\delta\hat\mu_i\rangle|\}<U$. Here, the transverse laser determines the transverse density distribution even when the cavity field is zero. In this case the MI lobes at $t=0$ shrink due to the classical incommensurate field. Due to the classical potential, in general no direct MI-SF transition is expected \cite{Pollet}. Figure~\ref{Fig:PhaseDiag}(a) displays  the order parameter, $\sum_i\langle \hat b_i\rangle/K$, in the $\mu-t$ plane and for density $\bar n \le 1$ \cite{Footnote:Bose-Glass}. The solid curve indicates where the gap in the spectrum is different from zero, corresponding to vanishing density fluctuations $\Delta\varrho=(\overline{n^2}-\overline{n}^2)^{1/2}$, for $\overline{n^2}=\sum_i \langle \hat{n}_i^2\rangle/K$ \cite{Lacki,Damski_06}. The dashed line separates the parameter region where the number of intracavity photons is at least 2 orders of magnitude larger than outside. For comparison the diagram for $n_{\rm cav}=0$ is reported in the inset. A larger region with vanishing order parameter appears when the atoms are strongly coupled to the cavity. The analysis of the corresponding density distribution shows a distinctively behaviour. Panel (b) displays the local density for $\delta_c=-5\kappa$ for the parameters indicated by the point in (a): The coupling with the resonator induces the formation of clusters with checkerboard density distribution, where the atoms scatter in phase into the cavity mode. At the border of the clusters the density fluctuates, so to allow the fields scattered by each cluster to interfere constructively. The inset shows that the clustering disappears when cavity backaction is negligible. In this latter case the field at the cavity output is zero, while when the effect of cavity backaction is relevant the corresponding field at the cavity output (panel (c)) has a finite intensity for a finite range of values of $t$:  When the tunneling rate is instead sufficiently large, there is no clustering and the atomic density becomes uniform along the cavity axis. Thus, the competition between the mechanical effects of the cavity field and quantum fluctuations leads to the creation of these clusters, which exhibit properties ranging from a BG to a strongly-correlated SF, and then disappear when tunneling becomes large.  

These results shed new light onto the effect of cavity backaction on an atomic ordered structure. At ultralow temperatures, when the ordered medium does not support photon scattering into the cavity mode, cavity backaction forces the system into locally ordered clusters which are phase locked with one another, thereby increasing the intracavity photon number. This is an example of selforganization of the quantum gas that is triggered by the quantum fluctuations of the atomic motion.

The authors acknowledge discussions with A. Niederle and with G. De Chiara, S. Fern\'{a}ndez-Vidal, M. Lewenstein,  H. Ritsch,  and support by the European Commission (IP AQUTE), the European Regional Development Fund,  the Spanish Ministerio de Ciencia y Innovaci\'on (QOIT: Consolider-Ingenio 2010; QNLP: FIS2007-66944; FPI; FIS2008-01236; Juan de la Cierva), the Generalitat de Catalunya (SGR2009:00343), and the German Research Foundation.


\begin{thebibliography}{breitestes Label}
\bibitem{Footnote:1}
This generally takes place when the ratios $k/k_0$ is a rational number, for which one has to consider other quasi-resonant transitions of the atoms.
\bibitem{Footnote:2}
Term  $\delta \hat{t}_{i}=-V_1 J_{1}^{(i)}-\hbar\frac{s_0^2}{\hat{\delta}_{\rm eff}^2+\kappa^2}\hat{\Phi}\left( 2\hat{\delta}_{\rm eff}Z_{1}^{(i)}+u_0\hat{\Phi} Y_1^{(i)}\right)$ is such that $|\langle \delta \hat{t}_{i}\rangle|/t^{(0)} \sim 10^ {-8}$.
\bibitem{Footnote:Batrouni}
We use a grand-canonical realization of the QMC \cite{Batrouni:1}, which allows one to better tracking the discontinuity of $\bar{n}$ at the phase transition border of the BG phase (Fig 2(b)). For the MI-SF border, a conventional canonical simulation is used \cite{Batrouni:2}, which is more suitable to determine the onset of the incompressibility of the MI. The blue points in Fig.~\ref{Fig:MC}(a) are determined when the jumps in the plots of $\bar{n}$ versus $\mu$ occur.
\bibitem{Footnote:Bose-Glass}
For $\Delta_a<0$ the MI lobe with $\bar n=1$ at $t\to 0$ is in the interval $0\le \mu\le \mu_1$, with $\mu_1<1$, and the corresponding BG phase appears for rational values of the densities $1<\bar n<2$ corresponding to $\mu_1<\mu<1$.
\bibitem{Footnote:Defect}
Localied regions in (a) at smaller order parameters are found deep in the SF phases. They are also observed in the inset and are due to the transverse incommensurate pump.
\bibitem{Footnote:MFthreshold}
In the local mean-field approach the threshold for the phases boundaries can not be found exactly [A. Niederle, H. Rieger, in preparation]. Here the threshold for the order parameter and the density fluctuation are chosen at 0.02.
\end{thebibliography}
\end{document}